\documentclass{pasj00}
\begin{document}
\SetRunningHead{S.\ Yamauchi et al.}{Thermal and Non-thermal X-Rays from the 
Galactic Supernova Remnant G348.5$+$0.1}
\Received{}
\Accepted{}

\title{Thermal and Non-thermal X-Rays from the Galactic Supernova Remnant 
G348.5$+$0.1}

\author{Shigeo \textsc{Yamauchi}, Sari \textsc{Minami}, and Naomi \textsc{Ota},}
\affil{Department of Physics, Faculty of Science, Nara Women's University, 
Kitauoyanishi-machi, Nara 630-8506}
\email{yamauchi@cc.nara-wu.ac.jp}
\and
\author{Katsuji \textsc{Koyama}}
\affil{Department of Physics, Graduate School of Science, Kyoto University, 
Kitashirakawa-Oiwake-cho, Sakyo-ku, Kyoto 606-8502\\
Department of Earth and Space Science, Graduate School of Science, Osaka University, 
\\ 1-1 Machikaneyama-cho, Toyonaka, Osaka 560-0043}


\maketitle

\begin{abstract}
We report on Suzaku results of the two distinct regions in the 
Galactic supernova remnant G348.5$+$0.1: 
extended thermal X-rays (''soft diffuse'') at the northeast region 
and non-thermal X-rays (CXOU J171419.8$-$383023) at the northwest region. 
The X-ray spectrum of the soft diffuse is fitted with neither an ionization equilibrium 
nor a non-equilibrium (ionizing) 
plasma model, leaving saw-teeth residuals in the 1.5--3 keV energy band. 
The residual structures can be produced when free electrons are  
recombined to the K-shells of highly ionized Mg and Si ions.  
In fact, the X-ray spectrum is nicely fitted with a recombination-dominant plasma 
model.  We propose a scenario that the plasma in nearly full ionized state at high 
temperature quickly changed to a recombining 
phase due to selective cooling of electrons to a lower temperature of $\sim$0.5 keV. 
The spectrum of CXOU J171419.8$-$383023 is well explained by a simple power-law model 
with a photon index of 1.9, nearly equal to the typical value of pulsar wind nebulae.
Since the distance is estimated to be the same as that of the soft diffuse, 
we infer that both the soft diffuse and CXOU J171419.8$-$383023 are associated with the same object, 
SNR G348.5$+$0.1.

\end{abstract}

\section{Introduction}
G348.5$+$0.1 (CTB 37A) is a supernova remnant (SNR) discovered in the radio 
band \citep{Clark1975}. 
The radio image shows a shell-like structure at the north and 
a break-out morphology extending to the south. 
The radio angular size 
and the spectral index are 15$'$ and  0.3, respectively (\cite{Green2009} and 
references therein).
Around this SNR, CO molecular clouds were found \citep{Reynoso2000} 
and OH masers were discovered in the molecular clouds 
\citep{Frail1996, Reynoso2000}. These suggest interactions of the SNR shock with the 
surrounding molecular clouds.
The distance was estimated to be 11.3 kpc \citep{Reynoso2000} or 6.3--9.5 kpc 
\citep{Tian2012}.

X-ray emission from G348.5$+$0.1 was first discovered in the ASCA Galactic Plane 
Survey (Yamauchi et al. 2002, 2008).
The X-rays with a large absorption column of interstellar matter are located within 
the radio shell \citep{Yamauchi2008}.
\citet{Aharonian2008} reported Chandra and XMM-Newton results that G348.5$+$0.1 
consists of two components; one is an extended 
thermal X-ray emission at the northeast part (here ''soft diffuse'') 
and the other is a non-thermal X-ray 
source with a small extent at the northwest part, named as CXOU J171419.8$-$383023. 
The Suzaku spectrum of the soft diffuse was fitted  
with a two-component model, a collisional ionization equilibrium (CIE) plasma with 
a temperature of 0.63 keV and a solar 
abundance and a power-law component with a photon index
 of 1.6 \citep{Sezer2011}.
Based on the spatial and spectral properties, a center-filled thermal X-ray emission within the radio shell, 
\citet{Sezer2011} noted that G348.5$+$0.1 is a new member of mixed-morphology (MM) SNRs.

\citet{Aharonian2008} discovered very high energy $\gamma$-rays (VHE-$\gamma$) at 
the position of G348.5$+$0.1 with High Energy 
Stereoscopic System (H.E.S.S.), named as HESS J1714$-$385. 
The spatial distribution of the VHE-$\gamma$ is consistent with that of the  
molecular clouds.
At the northeast part of the SNR, GeV $\gamma$-rays were found with the Fermi Large 
Area Telescope (LAT) \citep{Castro2010}.

\begin{figure*}
  \begin{center}
    \FigureFile(17cm,12cm){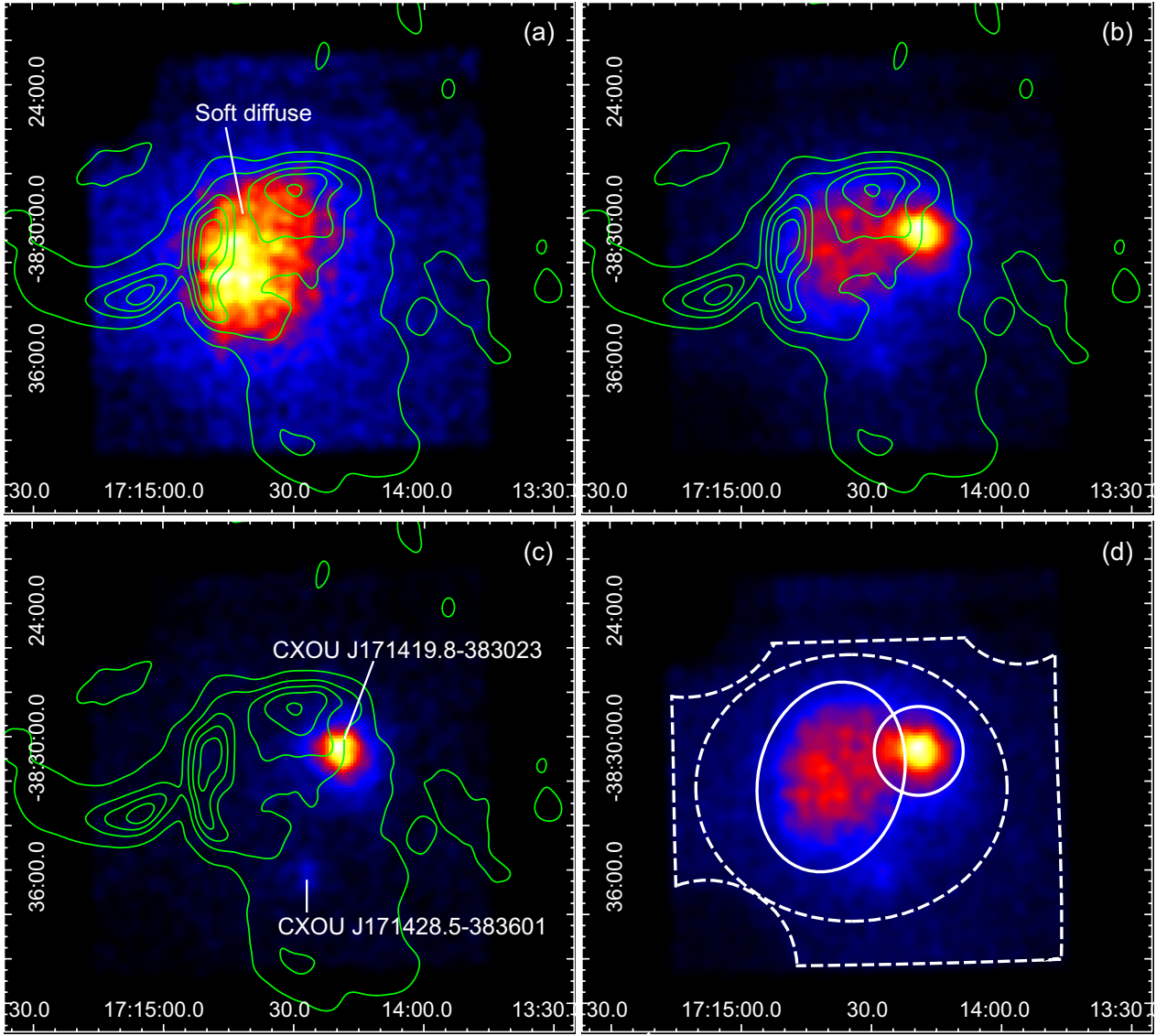}
  \end{center}
  \caption{
XIS images of G348.5$+$0.1 in the 0.7--2 (a), 2--5 (b), 5--8 (c), and 0.7--8 keV (d) 
energy bands (color). 
The coordinates are J2000.0. 
The radio map at 843 MHz using the Molonglo Observatory 
Synthesis Telescope (MOST) is displayed by the green contours in (a)--(c) 
\citep{Whiteoak1996}.
The X-ray images from XIS\,0, 1, and 3 were co-added, but 
neither the background subtraction nor the vignetting  correction was made. 
The images were smoothed 
with a Gaussian distribution with the kernel of $\sigma$=24$''$.
The intensity levels of the X-ray and radio bands are linearly spaced.
The solid ellipse and circle in (d) show the source regions for the soft diffuse
and CXOU J171419.8$-$383023, respectively,
while the background region is also shown by the dashed line in (d).   
}\label{fig:sample}
\end{figure*}

The X-ray spectra of most of the MM SNRs have been well explained with a thermal plasma 
in either an ionization equilibrium or an 
ionizing phase. Recently, on the other hand, a recombination-dominant plasma (RP) 
have been discovered from six MM SNRs 
\citep{Yamaguchi2009,Ozawa2009,Ohnishi2011, Sawada2012, Uchida2012, 
Yamauchi2013}. 
These SNRs share some common characteristics; 
all are accompanied with OH masers in surrounding molecular clouds and most of them are either 
GeV or TeV $\gamma$-ray sources. 

Although G348.5$+$0.1 shows the same characteristics, no RP has been reported 
so far (e.g., \cite{Aharonian2008, Sezer2011}),
possibly due to limited statistics \citep{Aharonian2008} or 
improper estimate of the large background in the Galactic ridge region \citep{Sezer2011}. 
Since Suzaku has the best performance for the spectral analysis of faint and diffuse 
sources, we reprocessed and reanalyzed the Suzaku data paying particular concern 
to background subtraction and a model fitting procedure. 
We found new results of 
possible evidence for the RP in the soft diffuse of the MM SNR G348.5$+$0.1 
and the most accurate value for physical parameters of the soft
diffuse  and 
CXOU J171419.8$-$383023.  In this paper, we report new revised results of the 
X-ray spectra and discuss the nature of G348.5$+$0.1.
Throughout this paper, the quoted errors are at the 90\% confidence level.

\section{Observation and Data Reduction}

Systematic survey observations studying various MM-SNRs 
on the Galactic plane were carried out with the Suzaku satellite \citep{Mitsuda2007}.
G348.5$+$0.1 was observed on 2010 February 20--21 (Obs. ID 504097010)
with the CCD cameras (XIS, \cite{Koyama2007}) 
placed at the focal planes of the thin foil X-ray Telescopes 
(XRT, \cite{Serlemitsos2007}). 
The pointing position was ($l$, $b$)=(\timeform{348D.44}, \timeform{+0D.09}). 

XIS sensor-1 (XIS\,1) is a back-side illuminated (BI) CCD, while XIS sensor-0, 2, 
and 3 (XIS\,0, 2, and 3) are front-side 
illuminated (FI) CCDs. The Field of view (FOV) of the XIS is 17$'$.8$\times$17$'$.8.
Since XIS\,2 turned dysfunctional in 2006 
November\footnote{http://www.astro.isas.jaxa.jp/suzaku/news/2006/1123/}, 
we used the data obtained with the other CCD cameras (XIS\,0, 1, and 3). 
A small fraction of the XIS0 area was not used, because of the data damage 
due possibly to an impact of micro-meteorite on 2009 
June 23\footnote{http://www.astro.isas.jaxa.jp/suzaku/news/2009/0702/}.
The XIS was operated in the normal clocking mode.
The spectral resolution of the XIS was degraded due to the radiation of cosmic 
particles 4.5 years after the launch.
It was restored by the spaced-row charge injection (SCI) technique.
Details of the SCI technique are given in \citet{Nakajima2008} and 
\citet{Uchiyama2009}.

Data reduction and analysis were made with the HEAsoft version 6.12 and
SPEX \citep{Kaastra1996} version 2.02.04. 
The XIS pulse-height data for each X-ray event were converted to 
Pulse Invariant (PI) channels using the {\tt xispi} software 
and the calibration database version 2012-07-03.
We rejected the data taken at the South Atlantic Anomaly, 
during the earth occultation, and at the low elevation angle 
from the earth rim of $<5^{\circ}$ (night earth) and $<20^{\circ}$ (day earth).  
The exposure time after these screenings was 53.8  ks. 

\section{Analysis and Results}

\begin{figure}
  \begin{center}
   \FigureFile(8cm,8cm){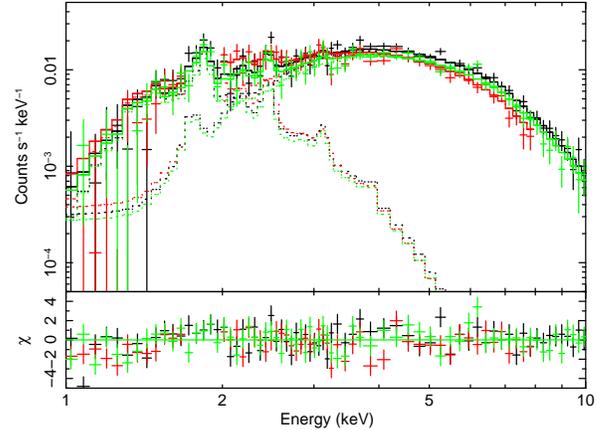}
  \end{center}
  \caption{
Upper: X-ray spectra of the CXOU J171419.8$-$383023 region (XIS0: black, XIS1: red, 
and XIS3: green) and the best-fit model.
Lower: residuals from the best-fit model.
 }\label{fig:sample}
\end{figure}

\begin{table}[t]
\caption{The best-fit parameters of CXOU J171419.8$-$383023 derived from a combined 
spectral analysis.}
\begin{center}
\begin{tabular}{lc} \hline  \\ [-6pt]
 Parameter & {Value} \\
\hline \\ [-6pt]
$N_{\rm H}$ ($\times10^{22}$ cm$^{-2}$)  & 7.6$^{+1.1}_{-1.0}$ \\
Photon index & 1.94$^{+0.15}_{-0.14}$  \\ 
\hline\\
\end{tabular}
\end{center}
\vspace{-10pt}
\end{table}

\subsection{X-Ray Image}

Figure 1 shows X-ray images of G348.5$+$0.1 in the 0.7--2, 2--5, 5--8, and 0.7--8 
keV energy bands.
To  maximize the photon statistics, the data of XIS0, 1, and 3 were combined.
We confirmed that the X-ray emission is located at the north part and found 
no significant X-ray emission 
in the break-out radio morphology extending to the south.

In the soft X-ray band (0.7--2 keV), diffuse X-rays are clearly seen at the 
northeast part of the SNR (soft diffuse).
On the other hand, we see a compact source at the northwest part in the hard X-ray 
band (CXOU J171419.8$-$383023, \cite{Aharonian2008}).
In addition to the soft diffuse and CXOU J171419.8$-$383023, 
another hard X-ray source was found at the south. 
It is identified with CXOU J171428.5$-$383601 \citep{Aharonian2008}.

\subsection{X-Ray Spectra}

In order to collect X-ray photons from the soft diffuse and CXOU J171419.8$-$383023
as many as possible, we extracted X-ray spectra from the partially overlapping two 
regions. The solid ellipse and circle in figure 1d show the source regions for 
the soft diffuse and CXOU J171419.8$-$383023, respectively.
The background data were taken from a source free region in the same FOV as large as possible,
which is shown by the dashed area in figure 1d.
In order to estimate the background counts in the source region carefully,
we should take account of the difference of vignetting effects between 
the source and background regions.
We made background-subtracted spectra by the following procedure as described in \citet{Hyodo2008}. \\
(1) We constructed the non-X-ray background (NXB) for the source and the background 
spectra from the night earth data using {\tt xisnxbgen} \citep{Tawa2008} and then 
subtracted the NXB from the source and the background spectra. \\
(2)The vignetting effect of the background spectrum was corrected 
by multiplying the effective area ratios between the source and
the background regions for each energy bin.\\
(3) Then, we subtracted the vignetting-corrected background data from the source region data.

Fractions of the background counts in the 1--8 keV band 
were 25--27\% and 18--22\% for the soft diffuse and the CXOU J171419.8$-$383023 region 
data, respectively.
Redistribution Matrix Files (RMFs) were made using {\tt xisrmfgen}, while  
Ancillary Response Files (ARFs), were simulated using  
a Chandra image for the photon distribution data in {\tt xissimarfgen}.
The XIS\,0, 1 and 3 spectra were simultaneously fitted.
Since the XIS gain calibration is known to be problematic around the Si K-edge energy,
we ignored the XIS1 data in the energy of 1.75--1.95 keV.

As we can see in figure 1d, the spectra extracted from the CXOU J171419.8$-$383023 region 
include some fraction of the soft diffuse X-rays, and vice versa. 
We therefore carried out combined fitting using a two-component model 
consisting of a thermal plasma (TP) and a power-law (PL) function. 
These models are based on the spatially resolved analysis with Chandra and XMM-Newton; 
the soft diffuse is fitted with a TP model, 
while CXOU J171419.8$-$383023 is fitted with a PL model
(\cite{Aharonian2008}, see also \cite{Sezer2011}).
In the combined fitting, we assume the spectral shape of the TP and PL
models are the same between the soft diffuse and the CXOU J171419.8$-$383023 regions.

\begin{figure}
  \begin{center}
   \FigureFile(8cm,16cm){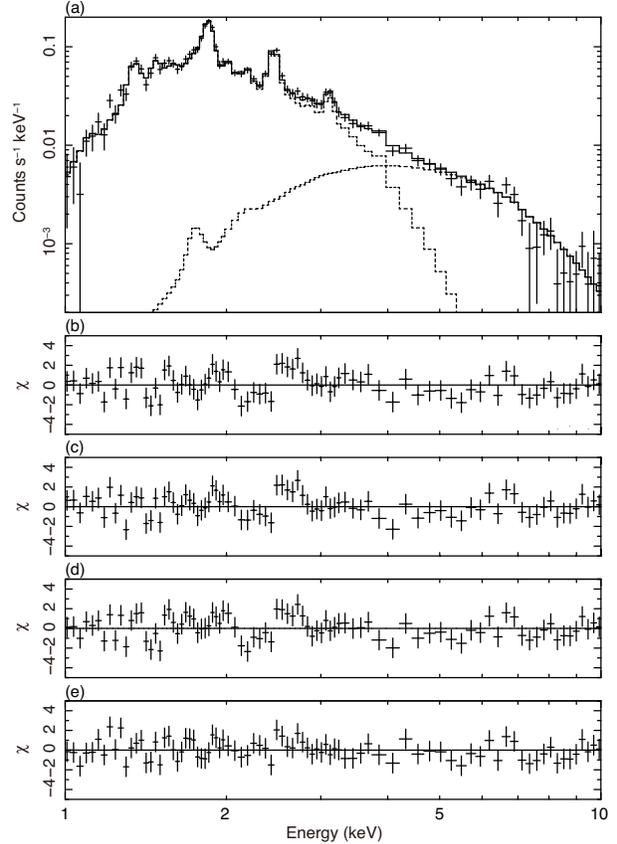}
  \end{center}
  \caption{
X-ray spectra (XIS0+XIS3) of the soft diffuse region and the residuals from the 
best-fit model. 
The histogram in (a) is the best-fit RP$+$PL model, while 
the residuals from the models of CIE$+$PL, 2CIE$+$PL, IP$+$PL, and RP$+$PL are shown
in (b), (c), (d) and (e), respectively (see table 2)
}\label{fig:sample}
\end{figure}

\begin{table*}[t]
\caption{The best-fit parameters derived from a spectral analysis for the soft 
diffuse region.}
\begin{center}
\begin{tabular}{lcccc} \hline  \\ [-6pt]
Parameter & \multicolumn{4}{c}{Value} \\
\hline \\ [-6pt]
Model   &  CIE$+$PL     & 2 CIE$+$PL        & IP$+$PL                 & RP$+$PL \\
\hline \\ [-6pt]
  &  XSPEC: {\tt vapec}   &  XSPEC: 2 {\tt vapec} & SPEX: {\tt neij} & SPEX: {\tt neij}\\
\hline \\ [-6pt]
$N_{\rm H}$ ($\times10^{22}$ cm$^{-2}$)  &
3.6$^{+0.1}_{-0.2}$ & 4.1$^{+0.4}_{-0.2}$ & 3.5$\pm$0.2 & 3.5$\pm$0.2\\
$kT_{\rm e\ [initial]}$ (keV)  & --- & --- & 0.001 (fixed) & 5 (fixed)\\
$kT_{\rm e}$ (keV)  & 0.63$^{+0.03}_{-0.02}$ & 0.70$^{+0.03}_{-0.05}$ & 
0.67$^{+0.06}_{-0.04}$ & 0.49$^{+0.09}_{-0.06}$\\
  &   & 0.17$^{+0.02}_{-0.04}$  &   &  \\
$n_{\rm e}t^{\ast}$ ($\times10^{12}$ cm$^{-3}$ s) & 
 --- & --- & 0.6$^{+0.8}_{-0.3}$ & 1.3$^{+0.3}_{-0.1}$\\
Mg$^{\dag}$= Al$^{\dag}$   & 
1.1$\pm$0.2 & 1.1$^{+0.4}_{-0.3}$ & 1.1$^{+0.4}_{-0.3}$ & 1.2$\pm$0.4 \\
Si$^{\dag}$   & 
1.0$^{+0.1}_{-0.2}$ & 1.3$^{+0.2}_{-0.2}$ & 1.3$^{+0.2}_{-0.3}$  & 1.6$\pm$0.3\\
S$^{\dag}$     & 
1.1$^{+0.1}_{-0.2}$ & 1.1$^{+0.2}_{-0.2}$ & 1.2$^{+0.2}_{-0.3}$  & 1.7$^{+0.3}_{-0.4}$\\
Ar$^{\dag}$ = Ca$^{\dag}$    & 
 1.9$\pm$0.5 & 1.4$^{+0.5}_{-0.4}$ & 1.6$\pm$0.7  & 3.1$^{+1.3}_{-1.4}$\\
Others$^{\dag}$  & 1 (fixed)  & 1 (fixed) & 1 (fixed) & 1 (fixed)\\
\hline \\ [-6pt] 
& \multicolumn{4}{c}{Power-law}\\
\hline \\ [-6pt]
$N_{\rm H}$ ($\times10^{22}$ cm$^{-2}$)  & 7.6 (fixed) & 7.6 (fixed) & 7.6 (fixed) & 7.6 (fixed)\\
Photon index & 1.94 (fixed) & 1.94 (fixed) & 1.94 (fixed) & 1.94 (fixed) \\ 
\hline \\ [-6pt] $\chi^2$/d.o.f. &
285/236 & 265/234 & 273/235 & 244/235 \\
\hline\\
\end{tabular}
\end{center}
\vspace{-10pt}
$^{\ast}$ Ionization or recombination timescale, 
where $n_{\rm e}$ is the electron density (cm$^{-3}$) and $t$ is the elapsed time 
(s).\\
$^{\dag}$ Relative to the solar value \citep{Anders1989}.\\
\end{table*}

Comparing the Suzaku X-ray image with the CO map \citep{Reynoso2000},
we found that the CO cloud covers the CXOU J171419.8$-$383023 region
but does not cover the most part of the soft diffuse region. We hence applied
independent absorptions ($N_{\rm H}$) for the TP and the PL
components.  
The applied model is

\smallskip
Absorption1$\times$TP$+$Absorption2$\times$PL.
\medskip

\noindent 
Here, we used a CIE ({\tt vapec} in xspec) model as a TP model.
The cross sections of photoelectric absorption were taken from 
Morrison and McCammon (1983)
and the abundance tables were taken from Anders and Grevesse (1989).
The abundances of Mg, Si, S, and Ar were free parameters, while 
those of Al and Ca were assumed to be the same as Mg and Ar, respectively. 
The other elements were fixed to the solar abundances.
Both the spectra of CXOU J171419.8$-$383023 and the soft diffuse regions were 
reasonably represented by this combined model with $\chi^2$/d.o.f. = 549/428.
The best-fit two-component model and the best-fit parameters of the PL model for 
CXOU J171419.8$-$383023 are shown in figure 2 and table 1, respectively.
As a TP contamination from the soft diffuse, we used a CIE model.
However, as we reveal in the following paragraphs,
a CIE model does not exactly represent the spectra of the soft diffuse (see figure 3 and table 2).
We therefore re-fitted the spectra from the CXOU J171419.8$-$383023 region with the PL model fixing 
the best-fit TP model given in table 2, and obtained the consistent PL parameters 
with those given in table 1.

For the soft diffuse spectra, we see significant residuals from the two-component model.
We therefore examined the soft diffuse spectra using several TP models
fixing the spectral parameters of the PL model (photon index and $N_{\rm H}$ value) 
to the best-fit values in table 1.  
At first, a CIE ({\tt vapec} in xspec) model was examined, 
but was rejected with $\chi^2$ value of 285 (d.o.f.=236).
The best-fit parameters are listed in table 2, while
the residuals are plotted in figure 3b. 
Although the fitting was simultaneously made for XIS0, XIS1 and XIS3, 
we show the co-added (XIS0+XIS3) results in figure 3. 
The best-fit thermal plasma parameters are consistent with those in \citet{Sezer2011}.
We, however, found clear residuals (saw-teeth structures) in the 1.5--3 keV energy band,
$\sim$10--40 \% excess above the model.
This excess is larger than the systematic error of the effective area in this energy band
estimated from the Crab spectra ($<$5\%)\footnote{http://www.astro.isas.jaxa.jp/suzaku/doc/suzakumemo/suzakumemo-2008-06.pdf}, and hence
the saw-teeth residuals are not instrumental but real.

We also tried to fit the spectra with a two-temperature CIE model with 
the same abundances for the two CIE components.
We found that the same structures remained in the residuals 
(the $\chi^2$ value of 265 for d.o.f.=234, figure 3c).

We next examined a TP model in a non-equilibrium ionization (NEI) state (ionizing 
plasma: IP, {\tt neij} in SPEX) 
with the assumption of an initial temperature of $kT_{\rm e\ [initial]}$=0.001 keV.  
Then, we confirmed that the model also gave the large $\chi^2$ value of 273 
(d.o.f.=235) and the residuals were essentially the same as those of the CIE model 
(figure 3d).
Since the leading edge energies of the saw-teeth structures at  $\sim$1.8 keV and  
$\sim$2.4 keV 
correspond the K-shell ionization energies of the He-like Mg and Si,
the residuals would be due to a radiative recombination continuum (RRC) which is a 
sign of the RP.
Thus, we examined the RP model ({\tt neij} in SPEX).
This model gave a smaller $\chi^2$ value than the IP model with $\Delta 
\chi^2$ = 29, 
but $kT_{\rm e\ [initial]}$ was not well constrained ($kT_{\rm e\ [initial]}>1.6$ keV). 
We then assumed $kT_{\rm e\ [initial]}$=5~keV as a physically reasonable value and tried 
the final {\tt neij} fitting. 
This model well represented the spectra with the $\chi^2$ value of 244 (d.o.f.=235).
The best-fit model and residuals are plotted in figure 3a and 3e, respectively.
No systematic residual is seen in the 1.5--3 keV band.  
The best-fit parameters are listed in table 2. 

\section{Discussion}

We reprocessed the Suzaku data of the Galactic SNR G348.5$+$0.1 paying particular concern to 
background subtraction. 
The reanalysis confirmed that G348.5$+$0.1 has two X-ray emission components, 
an extended soft X-ray emission at the northeast part dominant in the soft X-ray band 
(soft diffuse) and a hard X-ray source at the northwest part, CXOU J171419.8$-$383023.
We further obtained new and accurate information on the X-ray emissions.
In the following subsections, 
we discuss on the nature of the soft diffuse and CXOU J171419.8$-$383023.

\subsection{Soft Diffuse Emission}

The soft diffuse is located within the radio shell, and exhibited emission lines 
from highly ionized Mg, Si, S, Ar, and Ca, which clearly show that the X-ray 
emission is thermal origin.
Thus, we confirm that G348.5$+$0.1 is a member of MM SNRs, as noted by \citet{Sezer2011}.
The model fits for the soft diffuse with a CIE, a 2 CIE or a standard NEI (ionizing) plasma 
model gave strong saw-teeth residuals in the 1.5--3 keV band, 
while a RP model well represented the overall spectra.

The column density of the hydrogen atom ($N_{\rm H_{I}}$) along  the line-of-sight 
to G348.5$+$0.1 is (1.74--1.78)$\times10^{22}$ cm$^{-2}$ \citep{DL} or 
(1.29--1.43)$\times10^{22}$ cm$^{-2}$ \citep{LAB}.
Using the mean value and the deviation, we estimate 
$N_{\rm H_{I}}$=(1.6$\pm$0.3)$\times10^{22}$ cm$^{-2}$.
On the other hand, that of the hydrogen molecule ($N_{\rm H_2}$) is estimated to be 
(2.2$\pm$0.8)$\times$10$^{22}$ cm$^{-2}$ 
from the CO intensity ($W_{\rm CO}$) around G348.5$+$0.1 of 87--157 K km s$^{-1}$ and 
the conversion factor from CO to H$_2$ ($N_{\rm H_{2}}$/$W_{\rm CO}$) of 
(1.8$\pm$0.3)$\times$10$^{20}$ cm$^{-2}$ K$^{-1}$ km$^{-1}$ s \citep{H2}.
Then, 
the total column density is estimated to be
$N_{\rm H}$ = $N_{\rm H_{I}} +2 N_{\rm H_2}$  = (6.0$\pm$1.5)$\times$10$^{22}$ cm$^{-2}$.  
For the distance estimation of G348.5$+$0.1, we use the observed $N_{\rm H}$ of the 
soft diffuse because that of CXOU J171419.8$-$383023 
may have additional absorption by the foreground molecular clouds.  
The best-fit value of 3.5$\times10^{22}$ cm$^{-2}$ is 58\% of 
the total column density along the line-of-sight to G348.5$+$0.1.
If the line-of-sight length of the $N_{\rm H_I}$ and $N_{\rm H_2}$ estimation in the radio band
is assumed to be 17 kpc,
double the distance from the Sun to the Galactic center,
the distance of G348.5$+$0.1 is estimated to be 9.9$\pm$2.6 kpc.
This value is consistent with those of the previous estimation \citep{Reynoso2000,Tian2012}.
We therefore adopt the distance to be 10 kpc in the following discussion. 

The size of the soft diffuse plasma is $\sim8'\times6'$ ($\sim$23.3 pc$\times$17.5 pc at 10kpc).
Dividing a radius of the major axis (4$'$$\sim$11.6 pc) by a sound velocity 
at a temperature of 0.49 keV, we can estimate a dynamical age of the SNR to be 
2.4$\times$10$^{4}$ year.  
Assuming that the line-of-sight length of the plasma is equal to the minor axis
(the plasma shape is assumed to be an ellipsoid), the volume is estimated to be 
1.1$\times$10$^{59} f$ cm$^3$, where $f$ is a filling factor.
Adopting the best-fit volume emission measure of 8.9$\times$10$^{58}$ cm$^{-3}$
and $n_{\rm e}$=1.2$n_{\rm H}$, where $n_{\rm e}$ and $n_{\rm H}$ are the electron and 
hydrogen density, respectively, 
the mean hydrogen density is derived to be $n_{\rm H}$=0.82$f^{-0.5}$ cm$^{-3}$. 
Then the recombination time ($n_{\rm e}t$/$n_{\rm e}$) is 
calculated to be 4.2$\times$10$^{4}$$f^{0.5}$ years,  
nearly the same as the dynamical age, 
We thus propose that the plasma was full ionized at high temperature ($>$1.6 keV) 
in the early phase of 4.2$\times$10$^{4}$$f^{0.5}$ years ago, and then the plasma changed to 
a recombining phase due to selective cooling of electrons to a lower temperature of 
$\sim$0.5 keV. 

The RRC in 
the soft diffuse is weaker than those of the other RP-detected SNRs; 
the recombining timescale of the soft diffuse is 
1.3$^{+0.3}_{-0.1}\times10^{12}$ cm$^{-3}$ s, 
while those of the other RP SNRs are (0.1--0.7)$\times10^{12}$ cm$^{-3}$ s 
(\cite{Sawada2012,Uchida2012, Yamaguchi2012,Yamauchi2013}; S. Minami, private communication).  
Thus, the plasma in the soft diffuse of G348.5$+$0.1 would be the oldest among the RP SNRs. 

\subsection{CXOU J171419.8$-$383023}

The spectrum of CXOU J171419.8$-$383023 is well explained by a simple PL model 
with a photon index of 1.94$^{+0.15}_{-0.14}$ and a column density of 
$7.6^{+1.1}_{-1.0} \times 10^{22}$ cm$^{-2}$. 
These values are consistent with those with Chandra and XMM-Newton
within the errors \citep{Aharonian2008}, but our values are more accurate 
with smaller errors. 

The density of the northern cloud in the line-of-sight to CXOU J171419.8$-$383023 
was reported to be $n_{\rm H_2}$=660 cm$^{-3}$ \citep{Reynoso2000}.
This corresponds to $N_{\rm H}\sim4\times10^{22}$ cm$^{-2}$, 
assuming the uniform density along the line-of-sight length of 10 pc 
(roughly equal to the width of the northern cloud).
Since the best-fit $N_{\rm H}$ value of  CXOU J171419.8$-$383023 is larger than 
that of the soft diffuse by $N_{\rm H}$=4$\times10^{22}$ cm$^{-2}$,
it is likely that CXOU J171419.8$-$383023 is at the same distance of the soft diffuse (10 kpc), 
and hence we infer that  these two sources are physically associated with SNR G348.5$+$0.1.
The unabsorbed X-ray flux of CXOU J171419.8$-$383023 in the 0.5--10 keV band 
is calculated to be $1.1\times10^{-11}$ erg s$^{-1}$ cm$^{-2}$ 
which roughly agrees with the Chandra/XMM-Newton results. 

\citet{Aharonian2008} argued that CXOU J171419.8$-$383023 is
a candidate for a pulsar wind nebula (PWN), although their best-fit photon index 
is smaller than those of typical PWNe.
The weighted mean value of the photon index of 48 PWNe observed with Chandra is 2.01, 
in good agreement with our best-fit value, and hence CXOU J171419.8$-$383023 
becomes a more likely candidate of PWN. 

\citet{Possenti2002} and \citet{Mattana2009} presented the empirical relation 
between the X-ray luminosity and the characteristic age of pulsars in PWNe. 
Using the relation and the best-fit 2--10 keV band luminosity of 7$\times$10$^{34}$  erg s$^{-1}$ 
at 10 kpc, 
we can estimate the age of CXOU J171419.8$-$383023 
to be $\sim10^3$--$10^4$ yr, roughly consistent with the dynamical age of the soft 
diffuse plasma ($\sim2\times10^4$ yr).
The X-ray emission from CXOU J171419.8$-$383023 is extended by 
$\sim$\timeform{0.D05} \citep{Aharonian2008} ($\sim9$ pc at the distance of 10 kpc), 
consistent with those of middle-aged PWNe with characteristic ages of 
$\sim10^3$--$10^4$ yr ($\sim$2--10 pc, \cite{Bamba2010}). 
Thus many pieces of circumstantial evidence for PWN have been accumulating so far.
However, no pulsation has been found from CXOU J171419.8$-$383023.
We encourage deep pulsation search for the PWN candidate.

\bigskip

We would like to express our thanks to all of the Suzaku team. 
This work was supported by
the Japan Society for the Promotion of Science (JSPS); the Grant-in-Aid 
for Scientific Research (C) 21540234 (SY), 24540232 (SY), and 24540229 (KK),
Challenging Exploratory Research program  20654019 (KK), 
and Specially Promoted Research 23000004 (KK).


\end{document}